\documentclass[11pt, nofootinbib, notitlepage, pra, aps, tightenlines, onecolumn]{revtex4-2}
\usepackage{listings}
\usepackage{epsfig}
\usepackage{latexsym}
\usepackage[english]{babel}
\usepackage{enumerate}
\usepackage{amsmath, amssymb, graphicx, color}
\usepackage[lofdepth,lotdepth]{subfig}
\usepackage[colorlinks,citecolor=blue,linkcolor=blue]{hyperref}
\usepackage[normalem]{ulem}
\usepackage[justification=justified,singlelinecheck=false]{caption}

\newcommand{\beq}{\begin{equation}}
\newcommand{\beqs}{\begin{equation*}}
\newcommand{\eeq}{\end{equation}}
\newcommand{\eeqs}{\end{equation*}}

\def\tr{\textrm{tr}}

\newcommand\ket[1]{\vert #1\rangle}
\newcommand\bra[1]{\langle #1|}

\newcommand\ketbra[2]{| #1\rangle\langle #2|}

\begin{document}


\title{Predictive complexity of quantum subsystems}

\author{Curtis T.~Asplund}
\affiliation{Department of Physics \& Astronomy, San Jos\'{e} State University, San Jos\'{e} CA 95192-0106}

\email{curtis.asplund@sjsu.edu}

\author{Elisa Panciu}
\affiliation{Department of Physics, University of Maryland, College Park MD 20742-4111}

\begin{abstract}
We define predictive states and predictive complexity for quantum systems composed of distinct subsystems. This complexity is a generalization of entanglement entropy. It is inspired by the statistical or forecasting complexity of predictive state analysis of stochastic and complex systems theory, but is intrinsically quantum. Predictive states of a subsystem are formed by equivalence classes of state vectors in the exterior Hilbert space that effectively predict the same future behavior of that subsystem for some time. As an illustrative example, we present calculations in the dynamics of an isotropic Heisenberg model spin chain and show that, in comparison to the entanglement entropy, the predictive complexity better signifies dynamically important events, such as magnon collisions. It can also serve as a local order parameter that can distinguish long and short range entanglement.
\end{abstract}

\maketitle

\section{Introduction}
\label{sec:Intro}

Quantum systems of many particles are capable of an astonishing variety of complicated behaviors. Characterizing these behaviors in a useful way can be challenging.
Building on the foundation of entanglement entropy, and incorporating ideas from complex systems theory, we define new quantities for such systems: predictive states and predictive complexity. We then apply them to a prototypical example, a Heisenberg model spin chain, and find that they provide an improved measure of some dynamical processes and they can serve as promising new tools for analyzing general quantum dynamical systems. 

We are primarily guided by two lines of research. On one hand, entanglement entropy of lattice systems and field theories has been a highly successful tool in recent times, from the celebrated area law \cite{2010RvMP...82..277E} to many applications in quantum dynamics \cite{2006JSMTE..03..001D, 2016JSMTE..06.4002E}. On the other hand, we have the statistical or forecasting complexity of stochastic and complex systems \cite{1986IJTP...25..907G, PhysRevLett.63.105, Shalizi_2001}, which has developed into a sophisticated formalism that has been applied to numerous systems \cite{doi:10.1063/5.0062829, 2021Chaos..31h3114J, 2012NatPh...8...17C}. 

We distinguish our definition of predictive complexity from the quantum Hamiltonian complexity \cite{2012RPPh...75b2001O, 10.1561/0400000066}, which classifies systems according to complexity classes, and so does not determine a definite complexity value. It is also distinct from the quantum computational or gate complexity \cite{2009CMaPh.287...41A, 2017AnHP...18.3449B}, which is the basis for most work in high-energy physics and gauge/gravity (AdS/CFT) duality  \cite{Chapman:2021jbh, Caceres:2019pgf, 2024arXiv240317475M, 2024JHEP...08..241C}. In particular, our construction is information-theoretic and has no dependence on a choice of gates or a reference state or a particular basis.

Our definitions can be seen as quantum extensions of predictive state analyses of classical spin systems \cite{PhysRevE.55.R1239, 2003PhRvE..67e1104F, e19050214, e24091282}, cellular automata \cite{2004PhRvL..93k8701S}, various spatio-temporal dynamical systems \cite{2012arXiv1211.3760G, rupe2019disco} and to optimal quantum models of stochastic systems \cite{2012NatCo...3..762G, 2018PhRvL.120x0502B, 2018PhRvX...8a1025A, 2020PhRvL.125b0601L, 2020PhRvL.125z0501E, PhysRevA.99.062110}. It is similar in spirit to using the statistical complexity to analyze the distribution of measurement outcomes quantum systems \cite{2022JSP...187....4S, 2023JSP...190..106V, 2023arXiv230300162G}, and this study of the Heisenberg and other models \cite{2022Entrp..24.1161C}, but we do not involve measurement distributions of any particular observable nor, again, do we rely on the choice of a reference state. 

Several other quantities called ``complexity" have been applied to quantum and spin systems \cite{2009JMP....50l3528L, 2022PhRvA.105f2431B, 2020PNAS..11730241B, 2022npjQI...8...41S, 2022PhRvB.106c5143M}. These are defined differently to the complexity we define here, but relations between these would be interesting to investigate further.

\section{Methods and Definitions}
\label{sec:Def}

\subsection{Predictive equivalence}

Consider a quantum system composed of a subsystem $A$ and its complement or exterior $B$. We have in mind a finite set of interacting particles, spins, or qubits, with $A$ consisting of a subset. 
Accordingly, we write the finite-dimensional Hilbert space $\mathcal{H}$ of this system as 
\beq
\label{eq:factor1}
	\mathcal{H} = \mathcal{H}_A \otimes \mathcal{H}_B \ .
\eeq
For any density operator of such a system, one can form the reduced density operators using partial traces $\rho_A = \tr_B \rho$ and $\rho_B = \tr_A \rho$. In the case of a pure state, $\rho = \ketbra{\psi}{\psi}$, the von Neumann entropies are equal and give the entanglement entropy between $A$ and $B$:
\beq
\label{eq:equalEE}
	S_A = -\tr \left( \rho_A \log \rho_A \right)= -\tr\left( \rho_B \log \rho_B \right) = S_B \ . 
\eeq
From information theory (coding theorems), this entropy may be interpreted as the amount of information needed to describe the state of $A$, on average, using a suitable encoding of a basis of $\mathcal{H}_A$.
We may also interpret this as the amount of information about $B$, relevant to $A$, that has been lost due to tracing out $B$. 
This measures how much $A$ is entangled with $B$, and vice versa.
 
Grouping or coarse-graining outcomes of a stochastic process generally leads to a reduction in entropy, since less precise information is needed to describe the outcomes. Suppose some states in $\mathcal{H}_B$ were equivalent, in some sense, say in terms of how they affect the evolution of $\rho_A$.
Then we would expect to need less information about what's going on in $B$ to effectively predict the behavior in $A$. Below, we describe how to implement any such equivalence in terms of how it affects $\rho_A$ and the entropy. We focus on an equivalence defined by identical evolution of a subsystem for some time, and the consequently reduced entropy is thus a measure of the average minimal information needed to sufficiently predict the state of $A$, or observables in $A$, for that time. 
This is the rough idea for what we call the predictive complexity.

We are following the ideas of causal or predictive state analysis, sometimes also called computational mechanics \cite{1986IJTP...25..907G, PhysRevLett.63.105, Shalizi_2001}. In the context of classical and stochastic dynamical systems, this type of analysis defines an equivalence relation between states that yield equivalent probabilistic predictions for the system. 

In our quantum context, let us call two states $\ket{\phi_1}, \ket{\phi_2} \in \mathcal{H}_B$, equivalent if and only if they lead to the same dynamics in $A$. More precisely, for a given $\ket{\psi} \in \mathcal{H}_A$, we can form the states $\ket{\Psi_1} = \ket{\psi}\ket{\phi_1}, \ket{\Psi_1} = \ket{\psi}\ket{\phi_2} \in \mathcal{H}$ and the corresponding density operators $\rho_1 = \ket{\psi}\ket{\phi_1}\bra{\phi_1}\bra{\psi}$ and $\rho_2 = \ket{\psi}\ket{\phi_2}\bra{\phi_2}\bra{\psi}$. Let us write $\rho_1(t)$ and $\rho_2(t)$ for the time-evolved density operators, i.e., if $U(t) = U_t = e^{-iHt/\hbar}$ is the time-evolution operator for a time-independent, whole-system Hamiltonian $H$, $\rho_1(t) = U(t)\rho_1 U^\dagger(t)$. Then we can define $\ket{\phi_1}$ and $\ket{\phi_2}$ to be equivalent if and only if
\beq
\label{eq:equiv}
	\tr_B  \rho_1(t) = \tr_B \rho_2(t) \ ,
\eeq
for all $\ket\psi \in \mathcal{H}_A$ and for some interval of time, which we call the time horizon $t_\text{h}$.  

Reflexivity, symmetry and transitivity of the relation follow quickly from eq.~\eqref{eq:equiv}, so it is an equivalence relation. 
This also comes from it being defined by the equality of the images of a function on $\mathcal{H}_B$. 

It is also useful to express eq.~\eqref{eq:equiv} in terms of relative entropy, or Kullback-Leibler divergence, defined for any two density operators $\rho$ and $\sigma$ to be 
$S(\rho \Vert \sigma) \equiv \tr(\rho\log\rho - \rho\log\sigma)$.
Since the relative entropy of two density operators is zero if and only if they are equal, eq.~\eqref{eq:equiv} becomes
\beq
\label{eq:equiv2}
	S\left(\tr_B  \rho_1(t) \Vert \tr_B \rho_2(t)\right) = 0 \ .
\eeq
Equivalently, one could express this by saying the distance between the two density operators is zero for a given interval of time, under any metric on the space of density operators, such as the trace distance. In fact, all one needs to have well-defined equivalence classes is a partition of $\mathcal{H}_B$, and this could come from a generalization of eq.~\eqref{eq:equiv2} where the right hand side is not strictly zero, but just sufficiently close to zero so that it leads to sufficient clustering of states in $\mathcal{H}_B$ to be partitioned (in a way that respects the linear and inner product structure, as we describe further below). We plan to investigate such generalizations in future work. 

The relation clearly depends on the time horizon. In relativistic systems and in a case where $A$ is taken to be a compact spatial region, this would translate into a spatial horizon distance $r_\text{h} = t_\text{h}/c$. By relativistic causality, we expect all states of $\mathcal{H}_B$ that only differ in their configurations at distances from $A$ greater than $r_\text{h}$ to be equivalent to each other.
We have similar expectations for quantum systems obeying a Lieb-Robinson bound, since this similarly limits the speed of propagation of information \cite{lieb1972finite}. In that case, though, exponentially small corrections are allowed. This is where small corrections to the right hand side of eq.~\eqref{eq:equiv2} could be important. We are interested in the states in $\mathcal{H}_B$ that can be partitioned into equivalence classes in terms of their effect on the subsystem $A$. We will explicitly see how this works in the case of the Heisenberg model below.

As we will see, the linear and inner product structure of $\mathcal{H}$ combined with the unitary evolution operator naturally gives rise to equivalence classes that can be thought of as parallel affine subspaces, akin to superselection sectors. They may also be interpreted as projective subspaces in projective Hilbert space, which can be a useful perspective since it quotients out un-normalized state vectors and we are ultimately interested in normalized states and density operators.\footnote{Projective Hilbert space, or ray space, is the space formed from Hilbert space by taking the quotient by the relation $\ket\psi \sim z \ket\psi$, where $z$ is any non-zero complex number. It can be thought of as the space of physically distinct states in $\mathcal{H}$ \cite{Ashtekar:1997ud, 2021PhRvA.103f2218A, e26030225}.}

The predictive equivalence defined by eq.~\eqref{eq:equiv} is natural from the point of view of reduced density operators, but to define an equivalence relation with the desired linearity properties on the Hilbert space $\mathcal{H}_B$, we need a further requirement on the dynamics of the equivalent states. To see this, let $\ket{\phi_1}, \ket{\phi_2} \in \mathcal{H}_B$ be normalized, orthogonal state vectors satisfying eq.~\eqref{eq:equiv}. Then let $\ket{\phi_3} = a_1 \ket{\phi_1} + a_2 \ket{\phi_2} \neq 0$ with 
$|a_1|^2 + |a_2|^2 = 1$, i.e., an arbitrary normalized linear combination. 
The evolution of this linear combination can be written
\begin{align}
	U_t \ket{\psi}(a_1 \ket{\phi_1} + a_2 \ket{\phi_2}) &= a_1 c_1^{ij}(t) \ket{\psi^i}\ket{\phi^j} \nonumber \\
		&+ a_2 c_2^{ij}(t) \ket{\psi^i}\ket{\phi^j},
\end{align}
where $i$ and $j$ are summed over bases for $\mathcal{H}_A$ and $\mathcal{H}_B$, respectively. 
A short calculation shows that in order that the property in eq.~\eqref{eq:equiv} holds for $\rho_3$, we need, for all $i$ and $k$,
\beq
\label{eq:orth_B}
	\sum_j c_1^{ij}(t) c_2^{kj*}(t) = 0 \ .
\eeq
This can be understood as requiring that $U_t$ maintain orthogonality between equivalent states in $\mathcal{H}_B$, and it can also be rewritten
\beq
\label{eq:orth_B2}
	\tr_B  \ket{\Psi_1 (t)}\bra{\Psi_2 (t)} = 0 \ .
\eeq
It is straightforward to show that this holds when $A$ and $B$ are decoupled, since then we may write $\ket{\Psi_i (t)} = \ket{\psi (t)}\ket{\phi_i (t)}$ and thus 
\begin{align}
	&\tr_B  \ket{\Psi_1 (t)}\bra{\Psi_2 (t)} \nonumber \\
	&= \bra{\phi^i} \ket{\psi (t)}\ket{\phi_1 (t)}
		\bra{\psi (t)}\bra{\phi_2 (t)} \ket{\phi^i} \nonumber \\
	&= \bra{\phi_2 (t)}\ket{\phi^i}  \bra{\phi^i} \ket{\phi_1 (t)} \ket{\psi (t)}\bra{\psi (t)} \nonumber \\
	&= \langle \phi_2 (t) \vert \phi_1 (t)\rangle \ket{\psi (t)}\bra{\psi (t)} \ ,
\end{align}
which equals zero since $\ket{\phi_1}$ and $\ket{\phi_2}$ were assumed to be orthogonal, and this is preserved if the dynamics in $A$ and $B$ are decoupled.

To see what this means in terms of the system Hamiltonian $H$ that describes local interactions in a spatially extended system, let us assume that it can be written in the form $H = H_A + H_B + H_{AB}$, where all these terms are assumed to be time independent and where $H_A$ and $H_B$ are defined in terms of operators that only act on $\mathcal{H}_A$ and $\mathcal{H}_B$ respectively. In particular, $[H_A, H_B] = 0$ and $H_{AB}$ contains all terms that couple the subsystems $A$ and $B$. In general, $[H_A, H_{AB}] \neq 0$ and $[H_B, H_{AB}] \neq 0$, but we expect these commutators to be relatively small, since they only involve operators on the boundary between $A$ and $B$. More precisely, in a $d$-dimensional lattice system with $O(N^d)$ sites, we expect expectation values of $H_A$ and of $H_B$ to be $O(N^d)$, while those of $H_{AB}$ and its commutators to be $O(N^{d-1})$. 

This is important as we analyze the criteria in eq.~\eqref{eq:orth_B} and eq.~\eqref{eq:orth_B2}. We first note that orthogonality is preserved under the evolution generated by $H_B$ alone, i.e.,
\begin{equation}
\langle \phi_1 |\phi_2\rangle = \langle \phi_1| e^{iH_Bt} e^{-iH_Bt} |\phi_2\rangle = 0 \ .
\end{equation}
Now write the full evolution operator as $U_t = \exp[-i(H_A + H_B + H_{AB})t/\hbar]$.
By using the Zassenhaus formula (a relative of the Baker–Campbell–Hausdorff formula) twice, we can write
\begin{align}
\label{eq:Zassenhaus}
	U_t &\approx	e^{-iH_B t/\hbar} e^{-iH_A t/\hbar} e^{-iH_{AB} t/\hbar}  \nonumber \\
&\times e^{-\frac12 [H_A, H_{AB}] (-it/\hbar)^2} e^{-\frac12 [H_B, H_{AB}] (-it/\hbar)^2} 	\ , 
\end{align}
where higher order factors involve nested commutators of $H_B$ and $H_{AB}$, higher powers of $-it/\hbar$, and increasingly small coefficients in the exponents, all of which generally tends to suppress their relative importance \cite{2012CoPhC.183.2386C}. The factors leading to any violation of eq.~\eqref{eq:orth_B2} are thus systematically suppressed. This is in addition to the difference in scaling of $H_A$, $H_B$ and $H_{AB}$ already mentioned.
An exact analysis of these factors will depend on the system. 

For example, in this paper we look in detail at the the Heisenberg spin chain model with $N$ sites and periodic boundary conditions, which has Hamiltonian
\begin{equation}
	\label{eq:Hamil0}
	H = - J \sum_{n=1}^N \mathbf{S}_n \cdot \mathbf{S}_{n+1} \ ,
\end{equation}
where $\mathbf{S}_n = (S^x_n, S^y_n, S^z_n)$ is the spin operator at site $n$.
We take $A$ to be a block of adjacent spins going from site $1$ to $N_A$, and in this case
\begin{equation}
\label{eq:HAB}
	H_{AB} = -J(\mathbf{S}_{N} \cdot \mathbf{S}_{1} + \mathbf{S}_{N_A} \cdot \mathbf{S}_{N_A+1}) \ ,
\end{equation}
with $\mathbf{S}_n = (S_n^x, S_n^y, S_n^z)$.
From this we can see that $[H_A, H_{AB}]$ and $[H_B, H_{AB}]$ will involve terms of the form
\begin{align}
	[\mathbf{S}_{n-1} \cdot \mathbf{S}_{n}, &\mathbf{S}_{n} \cdot \mathbf{S}_{n+1}] =
		[S_{n-1}^i S^i_{n}, S^j_{n} S^j_{n+1}] \nonumber \\
		&= S_{n-1}^i S^j_{n+1} [S^i_{n}, S^j_{n}] \nonumber \\
		&= S_{n-1}^i S^j_{n+1} i\hbar\epsilon^{ijk} S^k_n \nonumber \\
		&= i\hbar\epsilon^{ijk} S_{n-1}^i S^j_{n+1} S^k_n \nonumber \\
		&= i\hbar (\mathbf{S}_{n-1} \times \mathbf{S}_{n+1})\cdot \mathbf{S}_n \ ,
\end{align}
where we have used the fact that spin operators at different sites commute. 
Applying this, we have
\begin{align}
\label{eq:com1}
	[H_B, H_{AB}]/i\hbar &= (\mathbf{S}_{N-1} \times \mathbf{S}_{1})\cdot \mathbf{S}_N \nonumber \\
	&- (\mathbf{S}_{N_A} \times \mathbf{S}_{N_A + 2})\cdot \mathbf{S}_{N_A + 1} \ ,
\end{align}
and 
\begin{align}
\label{eq:com2}
	[H_A, H_{AB}]/i\hbar &= (\mathbf{S}_{N_A - 1} \times \mathbf{S}_{N_A + 1})\cdot \mathbf{S}_{N_A} \nonumber \\
	&- (\mathbf{S}_{N} \times \mathbf{S}_{2})\cdot \mathbf{S}_{1} \ . 
\end{align}
We see from eqns.~\eqref{eq:HAB}, \eqref{eq:com1} and \eqref{eq:com2} that $H_{AB}$ and its commutators scale like the boundary of $A$, which in this one dimensional case is just fixed to be two lattice points, or $O(N^0)$ (while $H_A$ and $H_B$ will generally have $O(N)$ terms). When inserted into the Zassenhaus expansion in eq.~\eqref{eq:Zassenhaus}, these factors may lead to some violation of eq.~\eqref{eq:orth_B2}, but this will be suppressed in the limit of a large lattice. This is consistent with the Lieb-Robinson bound, which allows for small corrections \cite{lieb1972finite}. For this work, we will ignore these small corrections and assume that predictive equivalence extends to arbitrary linear combinations, but it would be interesting to study this further. 

%


\subsection{Predictive states}
\label{sec:Proj_op} 

Any equivalence relation on $\mathcal{H}_B$ can be seen as a map from it to a quotient space $\mathcal{H}'_B$, which we'll call the predictive state space. We will see below that the quotient map is a linear projection map and $\mathcal{H}'_B$ can be considered a Hilbert space itself. 
Here, we focus on how the quotient map affects the reduced density operator $\rho_A$ and its entropy, and how to interpret the results information-theoretically. 

Let us consider a Hilbert space $\mathcal{H}_B$ of arbitrary dimension $N_3$.
Suppose $\{|\varphi_i\rangle\}$ is an orthonormal basis for a linear subspace of equivalent states, of dimension $N_1$. Let $N_2$ be the dimension of the orthogonal complement of the subspace, so that $N_1 + N_2 = N_3$.
We define a linear projection operator $P$ that maps all difference vectors $\ket{\varphi_i} - \ket{\varphi_j}$ to zero. This projects the subspace onto the span of the normalized vector $\ket{\gamma} = (1/\sqrt{N_1}) \left(\sum_{i=1}^{N_1} |\varphi_i\rangle \right)$ and may be written as 
$P = \text{Id}_{N_2} + \ket\gamma\bra\gamma$.

To have a probabilistic interpretation of the states after projection, they must be normalized. 
In line with the idea of grouping equivalent states, we require that the probability of the predictive state equal the sum of all the probabilities in the equivalence class.
A general state $\ket\Psi \in \mathcal{H} = \mathcal{H}_A \otimes \mathcal{H}_B$ may 
be written as
\begin{equation}
\ket{\Psi} = \sum_{i=1}^{\text{dim}\mathcal{H}_A} \sum_{j=1}^{N_3} a_{ij}\ket{\psi_i}\ket{\phi_j}\ .
\end{equation}
The normalization of the equivalent states will have a non-trivial effect on terms of the form 
$a_1 \ket\psi_i\ket{\varphi_1} + \ldots + a_{N_1} \ket\psi_i\ket{\varphi_{N_1}}$, i.e., terms that include equivalent states. 
In our prescription, these terms will be mapped to the following, in the normalized state $\ket{\Psi'}$:
\begin{equation}
\label{eq:primedstate}
\sqrt{\sum_j |a_j|^2} \frac{\sum_j a_j}{\left\vert \sum_j a_j \right\vert} \ket{\psi_i} \ket{\gamma} \ ,
\end{equation}
where the sums are from $1$ to $N_1$.
The first factor, involving a square root, is the normalization necessary for a consistent probabilistic interpretation of the coefficients. The second factor is a unit complex number that comes from the projection by $P$. This normalizing of coefficients makes the overall map, from the normalized states in $\mathcal{H}_B$ to the normalized states in $\mathcal{H}'_B$, non-linear.

In the case of additional, distinct subspaces of equivalent states, the discussion above is iterated and more coefficients will be involved. 
The equivalent subspaces may be directly summed into an increasing sequence and so, in linear algebra terms, form a (partial) flag for $\mathcal{H}_B$.
This is the same kind of structure that appears when decomposing a Hilbert space into superselection sectors \cite{2007RvMP...79..555B}.
The relevant projection operator $P$ is then the identity on the subspace of inequivalent states plus a sum over projectors 
$\ketbra{\gamma_i}{\gamma_i}$ onto rays, one for each distinct subspace of equivalent states.
We omit the explicit general expressions, since we don't need them for our purposes here.

In the end, we're interested in a normalized state $\ket{\Psi'}$ in the predictive state space $\mathcal{H}'_B$, and we have implicitly 
defined it to be $P\ket\Psi$ after its coefficients are normalized according \eqref{eq:primedstate}. 
We may then define the predictive state density operator $\rho' = \ketbra{\Psi'}{\Psi'}$ and its partial traces 
$\rho'_B = \tr_A \rho'$ and $\rho'_A = \tr_{B'} \rho'$. Here $\tr_{B'}$ means a partial trace over the the predictive state space 
$\mathcal{H}'_B$. We work out explicit examples of these definitions in Sec.~\ref{sec:SmallSys}.
The predictive states are used to define the predictive complexity, which we do next.

\subsection{Predictive complexity}

We define the \emph{predictive complexity} $C_A$ of $A$ to be the von Neumann entropy
of the predictive state density operator defined above, $\rho'_A = \tr_{B'} \ketbra{\Psi'}{\Psi'}$,
\begin{equation}
	C_A = -\tr \left( \rho'_A \log \rho'_A \right) \ ,
\end{equation}
where $\ket{\Psi'}$ is the state formed by implementing predictive equivalence, as defined by eq.~\eqref{eq:equiv}, and normalized according to eq.~\eqref{eq:primedstate}.
We interpret $C_A$ as the entropy of $A$ due to tracing over $A$'s environment $B$, modulo predictively equivalent states in $B$. One could similarly define other measures of predictive complexity from other information measures of $\rho'_A$, such as the Renyi entropy, negativity, etc.. 

We conjecture that $C_A \leq S_A = -\tr \left( \rho_A \log \rho_A \right)$, i.e., that the predictive complexity is bounded above by the standard entanglement entropy, with equality if no states in $B$ are predictively equivalent. This is true in all cases we've examined.
Similarly, in practice, one may only know that a certain subspace of states in $\mathcal{H}_B$ is predictively equivalent for some interval of time. 
For example, states supported outside of some horizon around $A$. The true set of equivalencies may be greater, and we expect this would only decrease the resulting complexity. In that case, one computes an upper bound on the predictive complexity. 

Many additional questions may be asked about the predictive complexity $C_A$ and its properties.
We address some of these in the Discussion, but many are the subject of future research, and we turn now to an example calculation of this quantity in a model system.

\subsection{Predictive states and complexity in a small system}
\label{sec:SmallSys}

We present a low-dimensional example where we can work out the definitions of predictive state and predictive complexity explicitly. Let's consider $\mathcal{H}_A \cong \mathbb{C}^2$ and $\mathcal{H}_B \cong \mathbb{C}^3$. Let $\{|1\rangle_A, |2\rangle_A \}$ be an orthonormal basis for $\mathcal{H}_A$ and $\{|1\rangle_B, |2\rangle_B, |3\rangle_B \}$ be one for $\mathcal{H}_B$, and suppose that $|1\rangle_B \sim |2\rangle_B$, i.e., they are found to be equivalent under eq.~\eqref{eq:equiv}. 
Form the normalized sum and difference vectors $\ket\alpha :=  (1/\sqrt{2})( \ket{1}_B -\ket{2}_B)$ and $\ket\beta :=  (1/\sqrt{2})( \ket{1}_B + \ket{2}_B)$.
We form the projection operator $P = \ket{3}_B \bra{3}_B + \ket\beta\bra\beta$ on $\mathcal{H}_B$ that projects along $\ket\alpha$ and onto its orthogonal complement, the image of $P$. Call that image $\mathcal{H}'_B$, in this case a two (complex) dimensional subspace, spanned by $\ket{3}_B$ and $\ket\beta$. 

Let's look first at a general state in the $1-2$ subspace of $\mathcal{H}_B$, where we have
$P(a_1\ket{1}_B + a_2 \ket{2}_B) = \frac{1}{2} (a_1 + a_2) (\ket{1}_B + \ket{2}_B) = \frac{1}{\sqrt{2}} (a_1 + a_2) \ket\beta$ .
If we normalized this state, we would end up with 
\beq
\label{eq:norm1}
 \frac{a_1+ a_2}{|a_1 + a_2|} \ket\beta\ .
\eeq 
We note that if $a_1 + a_2 = 0$, the state is mapped to the zero vector and so lies in the kernel of $P$.
Such states are effectively excluded from the final state space and are concomitant of having state vectors that are physically equivalent in some sense. The same phenomenon occurs in quantization of gauge theories, as in the BRST formalism, and we plan to elaborate on this connection in future work. 

Next, take a general normalized state in $\mathcal{H}$, given by 
\begin{align}
\label{eq:GenState1}
|\psi\rangle &= a_1 \ket{11} + a_2 \ket{12} + a_3\ket{13} \\ &+ a_4\ket{21} + a_5\ket{22} + a_6\ket{23} \nonumber \ ,
\end{align}
where $\ket{ij} = \ket{i}_A \otimes \ket{j}_B$. Under $P$, this state is mapped to 
the un-normalized state
$P\ket{\psi} = \frac{1}{\sqrt{2}} (a_1 + a_2) \ket{1}_A \ket\beta + a_3 \ket{1}_A \ket{3}_B +  \frac{1}{\sqrt{2}} (a_4 + a_5)\ket{2}_A \ket\beta + a_6 \ket{2}_A \ket{3}_B $ .

To normalize this state, we focus on the terms that have been affected by the projection. 
For example, the contribution of the first two terms in eq.~\eqref{eq:GenState1} to the norm of $|\psi\rangle$ was 
$|a_1|^2 + |a_2|^2$. For a consistent probabilistic interpretation of the coefficients, we need that to be the contribution of the first term in the projected state. Thus, we define the normalized state
\begin{align}
\label{eq:NormState1}
\ket{\psi'} &= \\
&\sqrt{|a_1|^2 + |a_2|^2}\frac{a_1+ a_2}{|a_1 + a_2|} \ket{1}_A \ket\beta + a_3 \ket{1}_A \ket{3}_B \nonumber \\
	+  &\sqrt{|a_4|^2 + |a_5|^2}\frac{a_4+ a_5}{|a_4 + a_5|} \ket{2}_A \ket\beta + a_6 \ket{2}_A \ket{3}_B \nonumber
\end{align}

We're interested in how this affects the reduced density operator $\rho_A = \tr_B |\psi\rangle\langle\psi |$, whose matrix elements are defined by
$\rho_A = \sum_{ij} \rho_{Aij} \ket{i}_A \bra{j}_A$. 
Represented as a matrix in the given basis, the original reduced density operator is
\begin{align}
\label{eq:rhoAorig}
\rho_A = \begin{pmatrix}
|a_1|^2 + |a_2|^2 + |a_3|^2 & a_1 a_4^* + a_2 a_5^* + a_3 a_6^* \\
a_1^* a_4 + a_2^* a_5 + a_3^* a_6 & |a_4|^2 + |a_5|^2 + |a_6|^2 
\end{pmatrix} \ .
\end{align}
Under the action of the equivalence relation, we have a modified reduced density operator
\beq
\rho'_A = \tr_{B'} |\psi'\rangle\langle\psi' | = 
	\langle \beta|\psi'\rangle\langle\psi'|\beta\rangle 
	+ \langle 3|\psi'\rangle\langle\psi'| 3\rangle \ .
\eeq
Represented as a matrix, the diagonal elements (and thus the trace) are unchanged, but the upper off-diagonal element is now
\begin{align}
\label{eq:offdiag1}
	&\rho'_{A12} = a_3 a_6^* +\\ 
	&\sqrt{(|a_1|^2 + |a_2|^2)(|a_4|^2 + |a_5|^2)}\frac{(a_1 + a_2)(a^*_4 + a^*_5)}{|a_1 + a_2||a_4 + a_5|} \nonumber
\end{align}
The other off-diagonal element is the complex conjugate, since $\rho'_A$ is Hermitian. 
The modification of the off-diagonals affects the eigenvalues and hence the von Neumann entropy of $\rho'_A$, which we identify as the predictive complexity $C_A$. The explicit expression for $C_A$ in terms of the $a_i$ is long and not illuminating, so we omit it.

To give a concrete example of how the predicitive complexity will differ from the entanglement entropy, suppose that we have a state that is a Bell state involving entanglement between subsystem $A$ and the states $\ket{1}_B$ and $\ket{2}_B$:
\beq
	\ket{\psi} = \frac{1}{\sqrt{2}}(\ket{11} + \ket{22}) \ .
\eeq
This corresponds to $a_1 = a_5 = 1/\sqrt{2}$ with the rest of the $a_i$ equal to zero. The reduced density operator, eq.~\eqref{eq:rhoAorig}, is diagonal and the entanglement entropy is clearly $S_A = \log 2$. Under the equivalence relation, $\ket{1}_B$ and $\ket{2}_B$ are in the same predictive state, and according to eq.~\eqref{eq:offdiag1} the reduced density operator $\rho'_A$ picks up off-diagonal elements equal to $a_1 a_5^* = 1/2$. This changes the eigenvalues to be $1$ and $0$ and so we get $C_A = 0$. This is consistent with the idea that in the case $\ket{1}_B \sim \ket{2}_B$, the state $\ket{\psi'} = (1/\sqrt{2})(\ket{1}_A \ket{\beta} + \ket{2}\ket{\beta})$ factorizes and the entanglement disappears. 

We have worked this out in a low-dimensional case, but the extension to an arbitrary finite-dimensional Hilbert space is relatively straightforward. In particular, in the following section, we will apply it to the $N = 32$ site Heinsenberg model.

\section{Results: Predictive complexity of the Heisenberg model}

\subsection{The Bethe ansatz}
\label{sec:Bethe}
It is interesting to apply these concepts to a quantum system with highly non-trivial spacetime dynamics, but that is nonetheless exactly solvable. 
We consider a standard spin chain from condensed matter theory, the 1D isotropic (XXX) Heisenberg model on $N = 32$ sites, with ferromagnetic, nearest neighbor interactions and periodic boundary conditions.
This model is defined by the following Hamiltonian:
\begin{equation}
	\label{eq:Hamil1}
	H = - J \sum_{n=1}^N \mathbf{S}_n \cdot \mathbf{S}_{n+1} \ ,
\end{equation}
where $\mathbf{S}_n = (S^x_n, S^y_n, S^z_n)$ is the spin operator at site $n$.
The value of $J$ does not affect our analysis, so we set $J=1$ (equivalently, we choose time $t$ to be in units of $\hbar/J$).  
Analyzing the Hilbert space of this model is greatly facilitated by the Bethe ansatz for the energy eigenstates, and we will largely follow
the very helpful review article \cite{doi:10.1063/1.4822511}. Within the two-magnon sector, we have calculated the complete time-dependent wavefunction and so can compute the entanglement entropies and predictive complexities exactly. This could also be done by direct diagonalization, but we found the Bethe ansatz helpful for distinguishing contributions from scattering states and bound states. 

The full Hilbert space has dimension $2^{32}$ and has an orthogonal basis of position states 
$\ket{\sigma_1 \ldots \sigma_N}$, the position basis, with $\sigma_n = \uparrow$ or $\downarrow$ representing spin up or down, respectively, in the $z$ direction at site $n$. These position states are generally not eigenstates of the Hamiltonian, and hence evolve non-trivially with time. We will restrict our attention to the two-magnon sector, a subspace of dimension $\binom{32}{2} = 496$, spanned by states $\ket{n_1,n_2}$, defined by spin flips at only sites $n_1$ and $n_2$. 

\begin{figure}[ht]
\centering
\subfloat[][]{\label{fig:EE1}\includegraphics[width=\columnwidth]{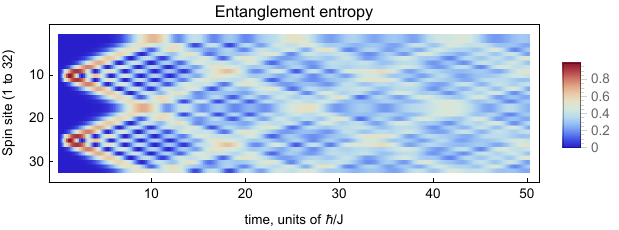}}\\
\subfloat[][]{\label{fig:complexity1}\includegraphics[width=\columnwidth]{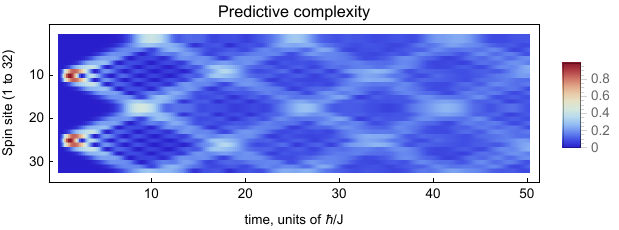}}
\caption[two-figures]{Spacetime plots of after two initial spin flips at sites 10 and 25. Fig.~\subref{fig:EE1} shows the single-site entanglement entropy, while Fig.~\subref{fig:complexity1} shows the single-site predictive complexity, with a horizon radius of two sites. The color scale indicates the value in bits (color map choice informed by \cite{moreland2009diverging}). Note the magnon beams and collisions appear with greater contrast in the complexity plot. We quantify this below. Calculations done with a time step of $dt = 0.2$.}
\label{fig:spacetimeplots}
\end{figure}


\subsection{Single site entropy and complexity}
\label{sec:Single_site}

The entanglement entropy of subsystems of various Heisenberg models and spin chains in various states has been studied by many authors, e.g., \cite{2014JSMTE..10..029M, 2022JHEP...02..072Z, 2021JPhA...54P4002E} and for a review see \cite{2016PhR...646....1L}. Here, we compare the entropy to the complexity for a subsystem $A$ consisting of a single spin, in a two-magnon state with an initial condition consisting of a largely ferromagnetic state but with two spins flipped. These states evolve in time and generate entanglement that spreads to all sites, as shown in the plot of the single site entanglement entropy in Fig.~\ref{fig:EE1}.

The predictive complexity for a given site $j$ may be defined for any choice of horizon radius, $r_\text{h}$. That is, the exterior region $B$ is defined to be all sites except site $j$, and we apply an equivalence relation on states in $B$ that only differ outside the horizon. 
This equivalence is straightforward to implement in the position basis. Effectively, we focus on the local environment of spin $j$, sites from $j - r_\text{h}$ to $j + r_\text{h}$ (modulo periodic boundary conditions). This is all that you need to predict the future of that spin for some duration of time, due to the Lieb-Robinson bound on the speed of propagation in this system \cite{lieb1972finite, 2004PhRvL..93n0402H, Chen:2023fiq}. This bound is associated with light-cone like dynamics, which can be seen in Fig.~\ref{fig:complexity1}.  

In more detail, we consider the case where $A$ consists of a single spin at some arbitrary site $j$ and the initial condition consists of two spins flipped. States whose only excitations (spin flips) are outside the horizon of $A$ are taken to be equivalent. The horizon is determined by the horizon radius $r_\text{h}$, which is a free (discrete) parameter. Because the dynamics of the Heisenberg model is local and obeys a Lieb-Robinson bound, the horizon distance is proportional to the predictive (time) horizon (we don't need the value of the Lieb-Robinson velocity for our purposes here, but calculating such values is an active area of research \cite{Chen:2023fiq, Wang:2019jsj}).  

As we mentioned above, the two-magnon sector has dimension $\binom{32}{2} =  496$. The positional basis for this sector consists of states of the form
\begin{align}
\label{eq:pos-basis}
	\ket{n_1,n_2} = |\uparrow\uparrow\cdots &\downarrow\cdots &&\downarrow\cdots\uparrow\uparrow\rangle \ ,  \\
	&n_1 &&n_2  \nonumber
\end{align}
where $1 \leq n_1 < n_2 \leq 32$, and it is orthonormal. 
The Bethe ansatz basis is composed of states of the form
\begin{align}
	\ket{k_1,k_2} &= \sum_{n_1 < n_2} a_{k_1 k_2}(n_1,n_2)\ \ket{n_1,n_2} \\
	\label{eq:kstate1}
	&= A_{k_1 k_2} \sum_{n_1 < n_2} {\big(} e^{i(k_1 n_1 + k_2 n_2 + \frac12 \theta)} + e^{i(k_1 n_2 + k_2 n_1 - \frac12 \theta)} {\big)} \ket{n_1,n_2} \ , 
\end{align}
where $A_{k_1 k_2}$ is a constant ensuring normalization: 
\beq
\langle k_1,k_2\mid k_1,k_2\rangle = \sum_{n_1 < n_2} |a_{k_1 k_2}(n_1,n_2)|^2 = 1 \ .
\eeq
Here $k_1$ and $k_2$ are lattice momenta and are related to the energy $E$ by 
\beq
\label{eq:Disp1}
E_{k_1 k_2} - E_0 = J (2 - \cos k_1 - \cos k_2) \ ,
\eeq
where $E_0$ is the energy of the state with no spin flips. 
The (scattering) phase $\theta$ is related to the momenta by
$2\cot (\theta/2) = \cot(k_1/2) - \cot(k_2/2)$. Translation invariance of the wavefunction imposes further conditions. With care and some numerical effort, a complete set of solutions may be calculated; see, e.g., \cite{doi:10.1063/1.4822511}. The spectrum includes a large class of states that may be interpreted as two-magnon scattering states, as well as bound states with complex momenta. 

As mentioned above, because of the locality of Hamiltonian, eq.~\eqref{eq:Hamil1}, and the Lieb-Robinson bound, if you want to predict the evolution of some segment of the spin chain for some amount of time, you only need information about the state of the system within a horizon region around $A$. Let us take $A$ to be the spin $\sigma_j$ at some arbitrary site $j$. For definiteness, take the horizon radius $r_\text{h} = 2$. We can depict these various regions on the spin chain as:
\begin{align}
\label{eq:chain1}
	\underbrace{\cdots \sigma_{j-3}}_{\text{outside}} \underbrace{\sigma_{j-2} \sigma_{j-1} \overbrace{\sigma_j}^A \sigma_{j+1} \sigma_{j+2}}_{\text{inside}} \underbrace{\sigma_{j+3}\cdots}_{\text{outside}} \ .
\end{align}
Recall that $B$ is defined to be the complement of $A$, i.e., all sites except $j$.

We are restricting to the two-magnon sector, so for a given value of $j$, all states $|n_1,n_2\rangle$ in $\mathcal{H}_B$ 
can be put into three types:
\begin{itemize}
\item I) $n_1$ and $n_2$ are outside the horizon.
\item II) Either $n_1$ or $n_2$ is inside the horizon, but not both.
\item III) Both $n_1$ and $n_2$ are inside the horizon.
\end{itemize}
These criteria can be phrased numerically, e.g., $n_1$ is inside the horizon if $|j - n_1 \mod N| \leq r_\text{h}$, etc. All states of type I are equivalent to each other. 
Thus, there are $N_\text{I} = \binom{N - (2r_\text{h}+1)}{2} = \binom{27}{2} = 351$ states in this equivalence class, in this case.
States in class II will be equivalent according to whether their inside spin flip is the same. So there are $2r_\text{h} = 4$ equivalence classes, with $N_\text{II} = N - (2r_\text{h}+1) = 27$ states in each class. Finally, each of the $\binom{2r_\text{h}}{2} =6$ states of type III are inequivalent to any other state. 

As described above, we can associate these equivalence relations with a map from the original Hilbert space to the space of (normalized) predictive states, see eq.~\eqref{eq:primedstate}.
In this case, we are interested in applying this equivalence relation (and ultimately calculating density operators and entanglement entropies) to dynamical, excited states, $|\Psi(t)\rangle$. In this work, we focus on states whose initial conditions are two individual spin flips. That is, states of the form:
\begin{equation}
\label{eq:Psi_t}
	|\Psi(t)\rangle = e^{-iHt} \ket{n'_1, n'_2}.
\end{equation}
The evolution of these two-magnon states can be expanded in terms of the Bethe ansatz states, which form a complete basis of Hamiltonian eigenstates within the two-magnon sector:
\begin{align}
	|\Psi(t)\rangle &= \sum_{k_1,k_2} e^{-iE_{k_1 k_2} t} \ket{k_1, k_2} \langle k_1, k_2|n'_1, n'_2\rangle \ \nonumber \\
\label{eq:psi_t}	
	&= \sum_{k_1,k_2} e^{-iE_{k_1 k_2} t}\ a^*_{k_1,k_2}(n'_1,n'_2)\ \ket{k_1, k_2} 
\end{align}
where the energies $E_{k_1 k_2}$ are given by eq.~\eqref{eq:Disp1}.
It will be useful to expand the energy eigenstates $\ket{k_1, k_2}$, in the position basis:
\begin{align}
\label{eq:psi_t2}
	\ket{\Psi(t)} &= \sum_{k_1,k_2} {\Bigg [}e^{-iE_{k_1 k_2} t}\ a^*_{k_1,k_2}(n'_1,n'_2)\times\left(\sum_{n_1 < n_2} a_{k_1,k_2}(n_1,n_2)\ket{n_1,n_2} \right){\Bigg ]} \ , 
\end{align}
and to define the time dependent matrix elements
\begin{align}
\label{eq:def_b}
	b(n_1, n_2, t) &= \langle n_1, n_2 \vert \Psi(t) \rangle \\
		&= \sum_{k_1,k_2} {\Bigg [}e^{-iE_{k_1 k_2} t}\ a^*_{k_1,k_2}(n'_1,n'_2)\times\left(\sum_{n_1 < n_2} a_{k_1,k_2}(n_1,n_2) \right){\Bigg ]} \ , \nonumber
\end{align}

To map the state in eq.~\eqref{eq:psi_t2} to its corresponding predictive state, apply the classification into types I, II \& III, above, to the sum over $n_1 < n_2$, then apply the formula in eq.~\eqref{eq:primedstate}. The result can be summarized as
\begin{align}
\label{eq:Psi_prime}
	&\ket{\Psi'(t)} = \sum_{k_1,k_2} {\Bigg [}e^{-iE_{k_1 k_2} t}\ a^*_{k_1,k_2}(n'_1,n'_2)\times \\
	&{\Bigg (} {\bigg(}\sum_{n_1, n_2 \in \text{out}} |a_{k_1,k_2}(n_1,n_2)|^2 {\bigg)}^{1/2} e^{i\theta_I} \ket{\gamma_I} + \nonumber \\
	&\sum_{n_1 \in \text{in}}{\bigg(}\sum_{n_2 \in \text{out}} |a_{k_1,k_2}(n_1,n_2)|^2 {\bigg)}^{1/2} e^{i \theta_{n_1}} \ket{\gamma_{n_1}} + \nonumber \\
	&\sum_{n_1,n_2 \in \text{in}} a_{k_1,k_2}(n_1,n_2)\ket{n_1,n_2} {\Bigg )} {\Bigg ]} \ . \nonumber
\end{align}
Here, the gamma states are those used to define the linear projection, specifically $\ket{\gamma_I} = (1/\sqrt{N_\text{I}}) \sum_{n_1, n_2 \in \text{out}} \ket{n_1,n_2}$ 
and $\ket{\gamma_{n_1}} = (1/\sqrt{N_\text{II}}) \sum_{n_2 \in \text{out}} \ket{n_1,n_2}$. 
The phase factors are normalized sums of the amplitudes, e.g.,
\begin{equation}
	e^{i \theta_I} = \frac{\sum_{n_1, n_2 \in \text{out}} a_{k_1,k_2}(n_1,n_2)}{\left| \sum_{n_1, n_2 \in \text{out}} a_{k_1,k_2}(n_1,n_2) \right|}
\end{equation}
The last three lines of \eqref{eq:Psi_prime} correspond to the classes I, II, and III, respectively. 

To compute entropies, one needs the matrix elements of the density operator and reduced density operator. These are determined fairly straightforwardly from the coefficients in \eqref{eq:Psi_prime}. For example, in main part of the paper we examined the single site entanglement entropy and predictive complexity, where the subsystem $A$ consisted of a single spin, and we set the horizon radius $r_\text{h} = 2$. The usual reduced density operator is then formed by tracing over the Hilbert spaces for all the sites except $A$. Similarly, the predictive complexity is computed from the modified reduced density operator $\rho'_A$, which is obtained by tracing over the predictive state space in the complement, $\mathcal{H}'_B$. It works out that the diagonal elements of $\rho_A$ and $\rho'_A$ are the same (as in the 5D Hilbert space example considered above), so let us turn to the off-diagonal elements. 

We note first that the off-diagonal elements of the unmodified $\rho_A$ are zero, essentially because we are working within the two-magnon sector, which is preserved by time evolution. 
To see this, note that 
\begin{align}
\label{eq:offdiag2}
	\bra{\downarrow_j}\rho_A \ket{\uparrow_j} &= \bra{\downarrow_j} \text{Tr}_B(\ket{\Psi}\bra{\Psi}) \ket{\uparrow_j} \ ,
\end{align}
and recall that $\ket{\Psi}$ is the time evolution of the two-magnon state $\ket{n'_1, n'_2}$ (see eq.~\eqref{eq:Psi_t}). 
For the off-diagonal element in eq.~\eqref{eq:offdiag2} to be non-zero, one would need a term in $\sum_B \langle\phi_B \vert \Psi\rangle$ (where the sum is over all $\ket{\phi_B} \in \mathcal{H}_B$) that simultaneously had a non-zero matrix element with $\langle\downarrow_j\mid$ and with $\langle\uparrow_j\mid$. Such a term would need to come from a state $\ket{\phi_B}$ that is simultaneously in the one- and two-magnon sector of $\mathcal{H}_B$, which does not exist. 

However, for $\rho'_A$, it is possible for the off-diagonal elements to be non-zero, because one- and two-magnon states can be lumped together in the same equivalence class. This allows the argument given above to be evaded. Indeed, for this particular calculation, this is the reason that the predictive complexity is different from the standard entanglement entropy at all. The off-diagonal elements can be worked out along the lines of eq.~\eqref{eq:offdiag1}. Applying this, we can calculate the off-diagonal elements in terms of the wavefunction elements defined in eq.~\eqref{eq:def_b}:
\begin{align}
\label{eq:offdiag3}
	\langle \downarrow_j\vert\rho'_A \vert \uparrow_j\rangle &= 
	{\Bigg(} {\bigg(}\sum_{n_1, n_2 \in \text{out}} |b(n_1,n_2,t)|^2 {\bigg)}  \\
	&\ \ \ \ \ \ \times 	{\bigg(}\sum_{n_2 \in \text{out}} |b(j,n_2,t)|^2 {\bigg)}
	{\Bigg)}^{1/2} \nonumber \\
	&\times \frac{\sum_{n_1, n_2 \in \text{out}} b(n_1,n_2,t)}{{\big |}\sum_{n_1, n_2 \in \text{out}} b(n_1,n_2,t){\big |}} \nonumber \\
	&\times \frac{\sum_{n_2 \in \text{out}} b(j,n_2,t)}{{\big |}\sum_{n_2 \in \text{out}} b(j,n_2,t){\big |}} \ . \nonumber
\end{align}

These non-zero off-diagonal elements affect the eigenvalues of $\rho'_A$, and thus also affect its von Neumann entropy. We identify the latter as the predictive complexity of the subsytem $A$. We present numerical calculations and analysis of this quantity for the 32-site Heisenberg model, which we obtained using the above formulas, implemented in Mathematica ver.~13.1 (code available upon reasonable request). We note that the phase factors in lines three and four of eq.~\eqref{eq:offdiag3} do not affect the eigenvalues of the two-by-two density matrix $\rho'_A$, so they do not need to be calculated to just get the predictive complexity. This significantly speeds up the calculation.

\begin{figure}[ht]
	\includegraphics[width=0.75\columnwidth]{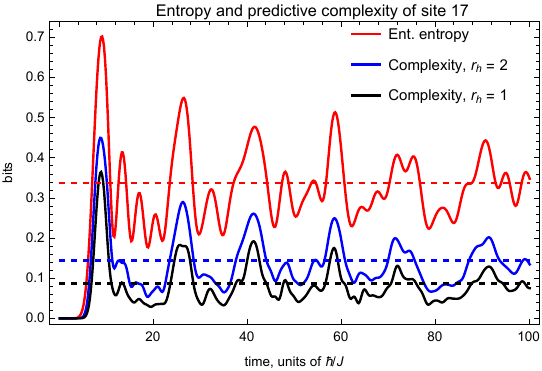}
	\caption{We show the evolution of the single site entanglement entropy and predictive complexity, for site 17, which sits in the middle of the two initial excited sites 10 and 25. The dashed horizontal lines indicate the equilibrium values. We note that the large local maxima in the three curves correspond to magnon collisions, and these peaks are more significant (relative to equilibrium) for the complexities than they are for the entanglement entropy. Calculations were done using a time step of $dt = 0.2$}
	\label{fig:ECplot}
\end{figure}

The single-site entanglement entropies and predictive complexities are shown in space-time diagrams in Fig.~\ref{fig:spacetimeplots}. 
We can see that the complexity appears to be more localized on the magnon beams and magnon collisions, while the entanglement entropy appears to show greater fluctuations throughout the spin chain. In particular, we can verify that the predictive complexity gives greater relative contrast to magnon collisions, in the sense that the local maxima associated with the collisions have a greater relative strength (compared to equilibrium), than appears with the entanglement entropy. This can be see in Fig.~\ref{fig:ECplot}, which plots the values for site 17 (midway between the initial spin flips), where the first magnon collision occurs at $t \approx 9.0$ (in units of $\hbar/J$). Quantitatively, the ratio of the first peak in the entanglement entropy to the equilibrium value, is $S_A(t=9.0)/\langle S_A\rangle \approx 2.08$. The corresponding ratio for the complexity, taking a horizon radius of one site ($r_\text{h} = 1$), is $C_A(t=9.0)/\langle C_A\rangle \approx 4.13$, or an increase of relative significance by a factor of almost two. Subsequent collisions are similar, though the increase is by a smaller factor.

The increased significance of the peaks, which we are interpreting as magnon collisions, can be understood from the fact that the single-site entanglement entropy is subject to contributions from entanglement between that single site and all other sites in the spin chain. The predictive complexity, on the other hand, discriminates between local and long-range entanglement. Since collisions are an inherently local phenomenon, it makes sense that the complexity would pick this up. The same reasoning can explain why fluctuations of the complexity are suppressed, compared with the entropy, as one can see in Fig.~\ref{fig:ECplot}. The standard deviation of $S_A(t)$ is approximately twice that of $C_A(t)$ (with $r_\text{h} = 1$) at late times, after the system has thermalized. These same effects are present, to a lesser extent, for larger values of the horizon radius.

\section{Discussion}
\label{sec:Discussion}

We have presented a generalization of the reduced density operator, based on the equivalence classes that we called predictive states, and defined its entropy to be the predictive complexity. When applied to the dynamics of the Heisenberg model, this complexity highlighted dynamically significant events like magnon propagation and, especially, collisions, with an improved effective signal-to-noise ratio. 
This indicates that it may be an improved local order parameter for some processes, and may have applications in magnon transport \cite{2023PhRvB.107r0403D} or quantum magnonics \cite{2022PhR...965....1Y}, and even more directly in recent experimental realizations of Heisenberg models \cite{2020Natur.588..403J, 2022PhRvA.106d3306P}.

Since the predictive complexity $C_A$ is inherently sensitive to short-range entanglement and since we expect $C_A \leq S_A$, the quantity $S_A - C_A$ may be a useful new measure of long-range entanglement. Relatedly, since $C_A$ depends on at least two length scales, the length scale of $A$ and the horizon radius $r_\text{h}$, we expect it to obey a variation of the usual area law for the entanglement entropy. Like the two-interval entanglement entropy (or mutual information) in conformal field theories \cite{Calabrese:2009ez}, it may be sensitive to non-universal information about the spectrum of states in the theory, but this is a subject for future work. 

We plan to study further properties of the predictive complexity in forthcoming work, including additivity properties and how it relates to other common definitions of complexity. The dynamics of other spin systems should also be interesting to study using these tools.

We also look to define predictive complexity for quantum field theories. This would allow us to make contact with many powerful results, such as for entanglement entropy in conformal field theory, and to compare with the symmetry-resolved entanglement entropy \cite{2019JPhA...52U5302B} and other generalizations of recent interest \cite{2022JHEP...05..152M}. 

\begin{acknowledgments}
C.A.~thanks Kassahun Betre for helpful discussions and feedback at various stages of the project, and Hilary Hurst, Ehsan Khatami, Eduardo Ibarra Garcia Padilla, and Nicholas Parilla for discussions and/or comments on drafts. We thank Sandra Chilson and Ileane Ho for their work on early ideas related to this project. We acknowledge the support of the Hackman Scholarship program of Franklin \& Marshall College, which supported E.P. during the summer when the initial ideas for this work were formulated. Research by C.A.~is supported by the U.S.~Department of Energy, Office of Science, Office of High Energy Physics RENEW-HEP program under Award Number DE-SC0024518.
\end{acknowledgments}

\bibliography{Complexity-refs6}{}
\bibliographystyle{apsrev4-1long}

\end{document}